\documentclass[a4paper]{jpconf}
\usepackage{epsfig}
\usepackage{amssymb}
\usepackage{rotating}
\usepackage{graphics}
\usepackage{graphicx}

\def\a{\alpha}

\def\d{\delta}

\def\h{\eta}

\def\k{\kappa}

\def\m{\mu}

\def\o{\omega}
\def\p{\pi}
\def\q{\theta}

\def\s{\sigma}
\def\t{\tau}

\def\O{\Omega}

\def\ve{\varepsilon}

\def\be{\begin{equation}}
\def\ee{\end{equation}}

\def\bea{\begin{eqnarray}}
\def\eea{\end{eqnarray}}

\begin{document}

\title{Dirac phase leptogenesis}

\author{Steve Blanchet}
\address{Max-Planck-Institut f\"{u}r Physik 
(Werner-Heisenberg-Institut)\\
 F\"{o}hringer Ring 6, 80805 M\"{u}nchen, Germany} 
\ead{blanchet@mppmu.mpg.de}

\begin{abstract}
I present here a concise summary of the preprint arXiv:0707.3024,
 written in 
collaboration with A. Anisimov and P. Di Bari. There we discuss leptogenesis 
when {\em CP} violation stems exlusively
from the Dirac phase in the PMNS mixing matrix. Under this assumption
it turns out that the situation is very constrained when a hierarchical
heavy right-handed (RH) neutrino spectrum is considered: 
the allowed regions are small and
the final asymmetry depends on the initial conditions. On the other
hand, for a quasi-degenerate spectrum of RH neutrinos, the {\em CP}
asymmetry can be enhanced and the situation becomes much more favorable,
 with no dependence on the initial conditions.
Interestingly, in the extreme case of resonant leptogenesis, in order
to match the observed baryon asymmetry of the Universe, we obtain a
lower bound on  $\sin \q_{13}$ which depends on the lightest active 
neutrino mass $m_1$. 

\end{abstract}


\section{Introduction}


With the understanding of the important role of flavor in
leptogenesis \cite{nardi,abada}, it has become progressively clear
that the link between 
low-energy {\em CP} violation and the size of the baryon asymmetry predicted
by thermal leptogenesis \cite{fy} is considerably tighter than in the
unflavored (or one-flavor) situation. In particular it was shown in
\cite{flavorlep} for a hierarchical RH neutrino spectrum that a 
non-zero Majorana phase can be the unique
source of {\em CP} violation contributing to the generation of the
baryon asymmetry of the Universe,
although the allowed region in the parameter space was small, and mainly
in the weak wash-out. The Dirac phase can also do the job, provided
the angle $\q_{13}$ is not too small \cite{pascoli}.

Within the see-saw model \cite{seesaw}, a nice way to see the different
contributions to the {\em CP} violation required for leptogenesis is to choose 
the following  parametrization of the Dirac mass matrix \cite{casas}:
\be
m_D=U\sqrt{D_m} \O\sqrt{D_M},
\ee
where $m_i$ and $M_i~(i=1,2,3)$ denote the light and heavy neutrino
masses, respectively, and $D_X\equiv {\rm diag}(X_1,X_2,X_3)$. 
Among the 18 supplementary parameters brought by the see-saw,
 6 are {\em CP} violating phases: 1 Dirac and 2 Majorana phases in the 
 PMNS mixing matrix $U$, as well as 3 unobservable phases in the 
$\O$ matrix. The $\O$ matrix can be conveniently parametrized as a 
product of three complex rotations,
$\O=R_{12}(\o_{21})R_{13}(\o_{31})R_{23}(\o_{32})$.

In \cite{dirac} we analysed in detail the role of the Dirac phase as the
{\em unique} source of {\em CP} violation required for leptogenesis. The aim is
here to summarize some important results obtained there.

\begin{center}
\includegraphics[width=0.31\textwidth]{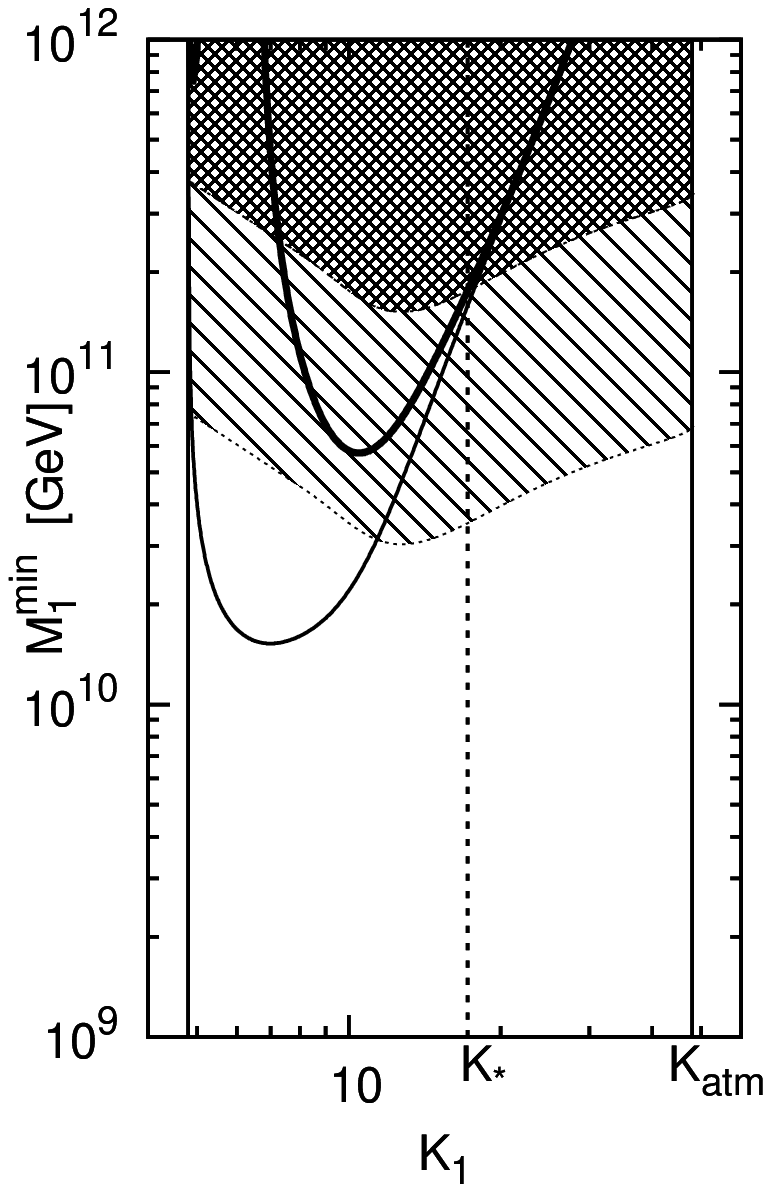}
\hspace{20mm}
\includegraphics[width=0.3\textwidth]{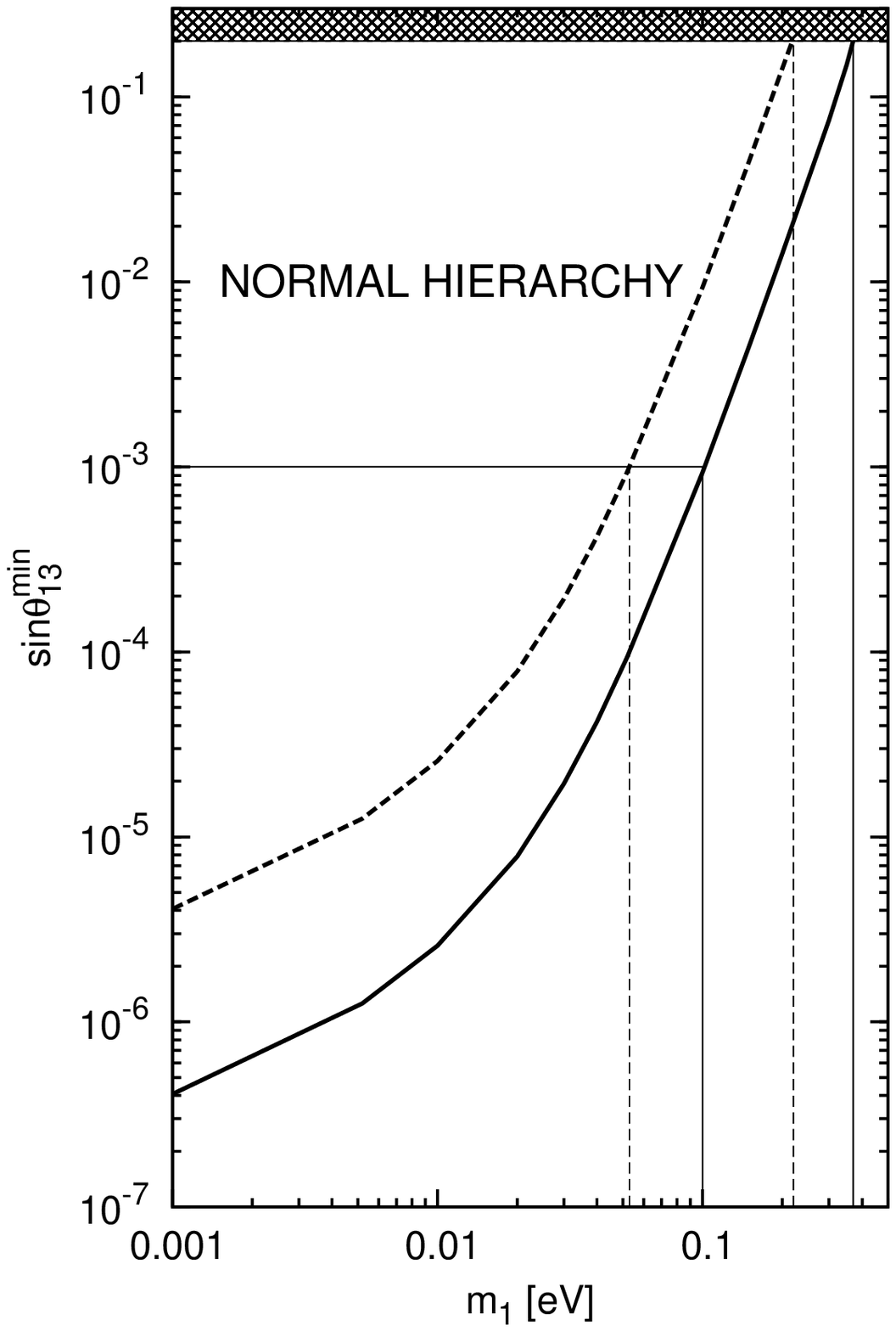}
\end{center}

\begin{tabular}[t!]{p{0.45\textwidth} p{0.45\textwidth}}
{\bf Figure 1.} HL, $\O=R_{13}$. Lower bound $M_1^{\rm min}$ vs. $K_1$ for
$m_1/m_{\rm atm}=0.1$, $s_{13}=0.2$, $\d=-\pi/2$ and
$\o_{31}<0$. The thin and thick solid lines denote thermal and vanishing
initial abundance of RH neutrinos, respectively.&
{\bf Figure 2.} DL, $\O=R_{13}$. Lower bound on
$\sin\theta_{13}$ vs. $m_1$ for $d=1$ (solid line) and
$d=10$ (short-dashed line).
Values $\sin\theta_{13} > 0.20$ are excluded at $3\,\s$ by current
experimental data.
\end{tabular}


\section{Hierarchical spectrum for the heavy RH neutrinos}


In the hierarchical limit (HL), $M_1\ll M_2\ll M_3$,
one can safely study only the decay of the lightest RH neutrino
\footnote{It was explicitly checked in \cite{dirac} that, for all cases under
study, the contribution from the heavier RH neutrinos was 
negligible.}, $N_1$,
and we anticipate that successful leptogenesis will occur for 
\mbox{$10^{9} {\rm GeV}\lesssim M_1\lesssim 10^{12} {\rm GeV}$}, 
where a two-flavor regime 
applies, with flavors denoted $\a=e+\m,\t$.

Since here we consider leptogenesis exclusively from one low-energy phase,
 namely the Dirac phase,
we have that the total {\em CP} asymmetry $\ve_{1}=\ve_{1\t}+\ve_{1,e+\m}=0$,
 so that the final asymmetry can be written as
\be
N_{B-L}^{\rm f}\simeq (\k_{1\t}^{\rm f}-\k_{1,e+\m}^{\rm f})\ve_{1\t},
\ee 
where $\k^{\rm f}_{1\a}$ are the final efficiency factors, which include 
the effects of the wash-out and the production of the asymmetry. 



Since the {\em CP} asymmetry parameter $\ve_{1\t}\propto M_1$, if one 
requires the final asymmetry, including the sphaleron conversion, to 
match the observed baryon asymmetry of
the Universe, $\h_{B}^{\rm CMB}=(6.1\pm 0.2)\times 10^{-10}$ \cite{WMAP3},
one obtains a lower bound on the lightest RH neutrino mass, $M_1$.

We show in Fig. 1 an example of lower bound $M_1^{\rm min}$ vs. 
$K_1\equiv \sum_k m_k |\O_{k1}|^2/m_{\star}~(m_{\star}\simeq 1.08\times
10^{-3}~{\rm eV})$ for the particular 
choice $\O=R_{13}$, with $m_1/m_{\rm atm}=0.1$, $\sin\q_{13}=0.2$ 
and $\d=-\p/2$. When the condition of validity for the fully flavored
Boltzmann equations is used (see \cite{zeno} for a discussion), the
region allowed is only outside the squared region. In the dashed region
it is not exluded but one expects some modification of the results shown.
It can be seen from Fig. 1 that the allowed region is very marginal,
 and only in the weak
wash-out regime, where there is a dependence on the initial conditions.
Moreover, in the example shown, we took the angle $\q_{13}$ at its 
maximal allowed value, so if the upper bound on $\q_{13}$ becomes one 
order of magnitude
more stringent in the future, it will essentially exclude Dirac phase
leptogenesis in the hierarchical limit \footnote{This rigorously applies
to the example shown in Fig. 1, but this statement turns out to apply
for all cases studied in \cite{dirac}.}.

One way-out of this unfavorable situation is to go to a quasi-degenerate
RH neutrino spectrum, where the {\em CP} asymmetry can be
enhanced and where a strong wash-out is ensured.


\section{Quasi-degenerate spectrum for the heavy RH neutrinos}


In the degenerate limit (DL), $M_1\simeq M_2 \simeq M_3$, we anticipate
the mass $M_1$ to be below $10^9~{\rm GeV}$, so that a three-flavor
problem arises, with flavors $\a=e,\m,\t$. Contrary to the HL, the {\em CP}
asymmetries, as well as the wash-outs, 
must be now summed over each RH neutrino:
\be
N_{B-L}^f\simeq \sum_{\a}(\ve_{1\a}+\ve_{2\a}+\ve_{3\a})\k_{\a}^{\rm f}
(K_{1\a}+K_{2\a}+K_{3\a}).
\ee
In the DL, the {\em CP} asymmetries are inversely proportional to 
the degeneracy parameter $\delta_{ji}\equiv (M_j-M_i)/ M_i$. This implies 
a dependence on $\d_{ji}$ of the lower bound on $M_1$ for successful
leptogenesis. For 
$\O=R_{13}$ we obtained the following $3\s$ bounds for normal and inverted
hierarchy, respectively:
\be
M_1\gtrsim 5.5\times10^9 {\rm GeV} {\d_{31}\over|\sin \q_{13}\sin \d|}\quad
{\rm and} \quad M_1\gtrsim 5\times10^{11} {\rm GeV} 
{\d_{31}\over|\sin \q_{13}\sin \d|}. 
\ee
It is interesting to notice how the degeneracy parameter is linked to the 
size of $\sin \q_{13}$ for successful leptogenesis at fixed $M_1$.

The enhancement for close RH neutrino masses is of course limited. 
Indeed, when 
$\d_{ij}^{\rm res}\simeq d M_i m_{\rm atm}/(16 \p v^2)$, one hits a resonance 
\cite{pilaftsis,abp}, where $d=1\div 10$ accounts
for the present uncertainty in the location of the resonance. It is interesting
to notice that in this extreme situation we obtained a lower bound on $\q_{13}$ 
which depends on the lightest active neutrino mass, $m_1$. This is shown
in Fig. 2, for $\O=R_{13}$ and a normal hierarchy. The corresponding
constraint for an inverted hierarchy as well as for other 
$\O$ matrices can be found in \cite{dirac}.
Note finally that for a vanishing $m_1$, successful Dirac phase leptogenesis
demands $\sin \q_{13}\gtrsim 2.3\times 10^{-7}$ and
$\sin \q_{13}\gtrsim 3\times 10^{-6}$ for normal and inverted hierarchy,
respectively.

\section{Conclusion} 

The possibility that the Dirac phase, which one hopes to measure in the 
next-generation neutrino experiments, might be the source of {\em CP} violation
required for leptogenesis is interesting. We obtained that a degenerate
RH neutrino spectrum is preferred and that in the extreme case of resonant
leptogenesis a nice link between low-energy parameters, such as $\sin \q_{13}$,
$m_1$ and the mass hierarchy, is present.

\ack

I would like to thank A. Anisimov and P. Di Bari for collaboration on this project.

\end{document}